\documentclass[12pt,notitlepage]{iopart}

\usepackage[numbers,sort&compress]{natbib}

\usepackage[colorlinks=true]{hyperref}

\expandafter\let\csname equation*\endcsname=\relax 
\expandafter\let\csname endequation*\endcsname=\relax

\usepackage{epsfig,amssymb,amsmath,mathrsfs,amsthm,graphicx,float,xcolor}

\usepackage{subfigure}

 \def\map#1{\mathcal #1}
\def\<{\langle}\def\>{\rangle}

\def\tr{\operatorname{tr}}
\def\St{\operatorname{St}}
\def\:{\hbox{\bf
    :}}

\def\R{\mathbb R}
\def\N{\mathbb N}
\def\C{\mathbb C}

\def\spc#1{\mathcal{#1}}

\def\set#1{\mathsf{#1}}

\newtheorem{theo}{{Theorem}}
\newtheorem{lemma}{{Lemma}}

\newtheorem{cor}{{Corollary}}
\newtheorem{remark}{{Remark}}
\newtheorem{defi}{{Definition}}

\begin{document}
\title{Energy requirement for implementing unitary gates on energy-unbounded systems}
 
\author{Yuxiang Yang}
\address{QICI Quantum Information and Computation Initiative, Department of Computer Science, The University of Hong Kong, Pokfulam Road, Hong Kong}

\author{Renato Renner}
\address{Institute for Theoretical Physics, ETH Z\"urich}

\author{Giulio Chiribella}
\address{QICI Quantum Information and Computation Initiative, Department of Computer Science, The University of Hong Kong, Pokfulam Road, Hong Kong}
\address{Department of Computer Science, Parks Road, University of Oxford, Oxford, OX1 3QD, UK}
\address{Perimeter Institute for Theoretical Physics, Waterloo, Ontario N2L 2Y5, Canada}

\begin{abstract} 
The processing of quantum information always has a cost in terms of physical resources such as energy or time. Determining the resource requirements is not only an indispensable step in the design of practical devices --- the resources need to be actually provided --- but may also yield fundamental constraints on the class of processes that are physically possible. 
Here we study how much energy is required to implement a desired unitary gate on a  quantum system  with a non-trivial energy spectrum. 
We derive a general lower bound on the energy requirement, extending the main result of Ref.~\cite{chiribella2021fundamental} from finite dimensional systems to systems with unbounded Hamiltonians. Such an extension has immediate applications in quantum information processing with optical systems, and allows us to provide bounds on the energy requirement of continuous variable quantum gates, such as displacement and squeezing gates.  
\end{abstract}
\maketitle

\medskip

\section{Introduction}
Determining the resource requirement of quantum information processing  is pivotal for its implementation. For this purpose, a wide range of quantum resource theories \cite{chitambar2019quantum} have been proposed and studied extensively, including coherence \cite{baumgratz2014quantifying,streltsov2017colloquium}, entanglement \cite{bennett1996mixed,horodecki2009quantum}, (a)symmetry \cite{marvian2012information,ahmadi2013wigner,tajima2018uncertainty,tajima2020coherence}, work \cite{sparaciari2017resource,faist2015minimal,faist2018fundamental}, and energy \cite{sagawa2009minimal,navascues2014energy,chiribella2017optimal,chiribella2021fundamental}.
In Ref.~\cite{chiribella2021fundamental}, the in-principle  energy requirement for a basic quantum information processing task, implementing a unitary gate on a finite dimensional system, has been determined.  
Specifically, consider a unitary operation $\map{U}$ on a system with Hamiltonian $H_S$ that one would like to implement with an error at most $\epsilon$. 
Then  any implementation requires a {\em battery} (i.e., an auxiliary system  serving as an energy supply), whose average energy  $\<  H_B\>$  is lower bounded as
\begin{align}\label{eq-finite-bound}
\<  H_B   \>  \ge  \frac{(\Delta E(\map{U})+\Delta E(\map{U}^{-1}))^2}{32\sqrt{\epsilon}\|H_S\|} \, ,
\end{align}
up to an error term that scales as $\sqrt{\epsilon}$.
Here $\|H_S\|$ denotes the operator norm of $H_S$ (as a convention, in this article we set the ground state's energy to zero), $\map{U}^{-1}$ denotes the inverse of $\map{U}$, and $\Delta E(\map{U})$ is the maximal gap between the input state energy and the corresponding output state energy of the unitary gate $\map{U}$, which captures the energy gain of the unitary.

 For finite dimensional systems, the bound (\ref{eq-finite-bound}) is achievable up to a constant (dimension-independent) factor, and it quantifies the minimum amount of energy resource needed by any concrete implementation. However, the bound does not work when the system has an unbounded Hamiltonian, in which case there are at least two obvious issues with Eq.~(\ref{eq-finite-bound}). First, for many common unitary gates acting on a quantum system with unbounded $H_S$, the energy gain $\Delta E(\map{U})$ depends on the energy of the input state, and larger input energies may thus correspond to larger energy gains. In these cases, $\Delta E(\map{U})$ can be infinite. For example, sending a coherent state $|\alpha\>$ through a displacement gate with displacement $\beta$, the output state is the coherent state $|\alpha+\beta\>$ (up to an irrelevant global phase), and the energy gain is $|\beta+\alpha|^2-|\alpha|^2$ that tends to infinity as $|\alpha|\to\infty$. In this case,   the bound (\ref{eq-finite-bound}) holds trivially (since the energy requirement for implementing the given gate {\em on all possible states} may indeed be infinite). However, it does not take into account that in many realistic scenarios the goal is not to implement a gate on every possible state, but rather on states satisfying a bound on the expectation value of the energy.  Another trickier issue is that $\|H_S\|=+\infty$ for unbounded Hamiltonians, which trivialises the bound.
Due to these issues, the existing bound (\ref{eq-finite-bound}) does not capture the energy requirement of gates that are common in quantum optics, including displacement \cite{glauber1963coherent}, squeezing \cite{yuen1976two}, and other non-linear operations.

In this article, we resolve both issues, obtaining the energy requirement for quantum processors acting on energy-unbounded systems. We circumvent the first issue (unbounded energy gain) by incorporating energy-constrained figures of merit as well as energy constraints on the processors \cite{winter2017energy,Shirokov2018,Gschwendtner2021infinite,pirandola2017ultimate,pirandola2017fundamental} into our framework. Under the energy-constrained scenario, the quantity $\Delta E(\map{U})$ in the previous bound (\ref{eq-finite-bound}) is replaced by a new term that depends on the input energy. Furthermore, we tackle the second issue (infinite norm of the Hamiltonian) by proposing an energy threshold method. By overcoming these two issues, our result extends the scope of Ref.~\cite{chiribella2021fundamental} to  a large class of continuous variable systems, with applications in quantum photonics as well as potential applications in fundamental physics (see Section \ref{sec-conclusion} for more details). 

The remaining part of the article is structured as follows. In Section \ref{sec-preliminary}, we introduce the basic notions as well as tools required for quantum information processing under energy constraints. In Section \ref{sec-mainresult}, we derive the energy requirement for implementing unitary gates on systems with unbounded Hamiltonians. In Section \ref{sec-applications}, we show how our new bound can be used to retrieve the main result of Ref.\ \cite{chiribella2021fundamental}, and we discuss applications to continuous-variable systems. Finally, in Section \ref{sec-conclusion}, we conclude the article with an overview of potential long-term applications and generalisations of our result.
 
\section{Preliminaries}\label{sec-preliminary}
\subsection{Basic notation}
In this article, we use the following notations. For a matrix $A$, positive-semidefiniteness is indicated by $A\ge0$.  
We denote by $\spc{H}$ a Hilbert space and by $\St(\spc{H})$ the set of density operators on $\spc{H}$. For a pure state $|\psi\>$, we adopt the notation $\psi:=|\psi\>\<\psi|$ for its density operator. Each system is associated with a Hamiltonian, denoted by $H$ (possibly with a subscript $S/B$ for the Hamiltonian of the system/battery, i.e., $H_S/H_B$). Here we assume $H$ to be grounded and discrete, i.e., $H=\sum_{n=0}^{\infty} e_n|e_n\>\<e_n|$ with $e_0\le e_1\le e_2\le\cdots$. For simplicity, we assume without loss of generality $e_0\ge0$. Fixing any system and its Hamiltonian, we write $P_{e'}:=\sum_{e\le e'}|e\>\<e|$ for the projection into its energy eigenspace with an energy threshold $e'$. For a quantum state $\rho$ on $\spc{H}$, we denote by $E(\rho):=\tr[\rho H]$ its energy.

A generic quantum process can be described by a completely-positive trace-preserving linear map from an input Hilbert space to a (possibly different) output Hilbert space, named a \emph{quantum channel}. Quantum channels that preserve the distance between quantum states are called \emph{isometries}. An isometry whose input and output spaces have the same dimension  is called a \emph{unitary} channel (or a {\em unitary gate}, or even just a {\em unitary}, for short). A unitary  $\map{U}$ acts on an input via the relation $\map{U}(\cdot)=U(\cdot)U^\dag$ for a unitary matrix $U$. Its inverse, denoted by $\map{U}^{-1}$, then follows the relation $\map{U}^{-1}(\cdot)=U^\dag(\cdot)U$.

\subsection{Energy-constrained metrics}
The similarity of two quantum channels $\map{A}$ and $\map{B}$ acting upon the same system $\spc{H}$ can be tested by sending a probe state $\Psi\in\St(\spc{H}_R\otimes\spc{H})$ (entangled to a reference register $R$ with $\spc{H}_R\simeq\spc{H}$) through the implementation and comparing the fidelity between the output state and the desired output:
\begin{align}\label{eq-statefid}
F_{\Psi}(\map{A},\map{B}):=F\left((\map{I}_R\otimes\map{A})(\Psi),(\map{I}_R\otimes\map{B})(\Psi)\right)
\end{align}
where $F(\rho,\sigma):=(\tr\sqrt{\sqrt{\rho}\sigma\sqrt{\rho}})^2$ is the quantum state fidelity. The similarity is evaluated via the channel fidelity, which equals the infimum of $F_{\Psi}$ over all possible probe states. 

For quantum channels acting on energy-unbounded systems, however, the conventional channel fidelity is ill-defined. For example, Winter  \cite{winter2017energy} pointed out that the (unconstrained) channel fidelity between any two quantum attenuators is always zero and thus fails to capture their true similarity. 

To address this issue, we adopt the energy-constrained fidelity \cite{Shirokov2018} as a figure of merit:
\begin{align}
F^E\left(\map{A},\map{B}\right):=\inf_{\substack{\Psi\in\St(\spc{H}_R\otimes\spc{H})\\\tr\Psi(I_R\otimes H)\le E}}F_{\Psi}(\map{A},\map{B})
\end{align}
with $F_\Psi$ given by Eq.~(\ref{eq-statefid}). Following the same idea, the diamond norm \cite{kitaev1997quantum} can be generalised to the energy-constrained worst-case error, introduced by Pirandola et al. \cite{pirandola2017fundamental}, Shirokov \cite{Shirokov2018}, and Winter \cite{winter2017energy}: 
\begin{align}
D^{E}(\map{A},\map{B}):=\sup_{\substack{\Psi\in\St(\spc{H}_R\otimes\spc{H})\\\tr\Psi(I_R\otimes H)\le E}}\frac12\|\map{I}_R\otimes(\map{A}-\map{B})(\Psi)\|_1
\end{align}
for any $E\in\R_+$. Here $\|\cdot\|_1$ denotes the trace norm (1-norm). 
When there are multiple input systems $\spc{H}_1\otimes\cdots\otimes\spc{H}_k$, it is often useful to put a constraint on the input energy of each individual subsystem rather than on the total input energy. For this purpose, the notion of energy-constrained worst-case error can be readily extended (and similarly for the energy-constrained fidelity) \cite{Gschwendtner2021infinite}: 
\begin{align}
D^{(E_1,\dots,E_k)}(\map{A},\map{B}):=\sup_{\substack{\Psi\in\St(\spc{H}_R\otimes\spc{H}_1\otimes\cdots\otimes\spc{H}_k)\\\tr\Psi H_i\le E_i\,\forall i\\H_i:=(I_R\otimes I_1\otimes\cdots H_i\cdots\otimes I_k)}}\frac12\|\map{I}_R\otimes(\map{A}-\map{B})(\Psi)\|_1
\end{align}
for any $(E_1,\dots,E_k)\in\R_+^k$.
The Fuchs-van der Graaf inequality \cite{fuchs1999cryptographic} can be extended to its energy-constrained version as:
\begin{align}\label{eq-fuchs}
1-\sqrt{F^E(\map{A},\map{B})}\le D^E(\map{A},\map{B})\le\sqrt{1-F^E(\map{A},\map{B})}.
\end{align}
In this way, the energy-constrained fidelity and the energy constrained diamond norm are related. High fidelity always implies low error and vice versa.

\subsection{Energy requirement of a unitary gate}
Here we define the task under consideration in this article, i.e., the  physical implementation (implementation, in short) of a given unitary using auxiliary systems and interactions that preserve the total energy. Consider a unitary gate $\map{U}(\cdot):=U(\cdot)U^\dag$ acting on a system $\spc{H}_S$ with grounded (but potentially unbounded) discrete Hamiltonian $H_S$. The goal is to determine the \emph{energy requirement} for the unitary $\map U$, i.e., how much energy is needed for its implementation. 
For the latter, we consider a model that features an explicit battery system $B$, which is initialised to a fixed state $\beta$. The battery provides or absorbs the energy consumed or released by the implementation. To ensure that no other energy is introduced into the system, we demand that the the time evolution $\map{V}(\cdot):=V(\cdot)V^\dag$ of the implementation conserves the total energy of the system and battery ($[V,H_S+H_B]=0$). We remark that explicit examples of such a time evolution $V$ for implementing a generic unitary $U$ can be found in Refs.~\cite{skrzypczyk2013extracting,navascues2014energy,aaberg2014catalytic,tajima2020coherence}, and an explicit form of the battery state for the case of $\|H_S\|<\infty$ is given in \cite[Section IV]{chiribella2021fundamental}.

We now demand that the evolution that $\map{V}$ induces on the system approximates the desired gate $\map{U}$:
\begin{defi}[$(E,\epsilon)$-ideal implementation]
An implementation $(\map{V},\beta)$ is said to be \emph{$(E,\epsilon)$-ideal} for some $\epsilon\ge0$ if $F^E(\map{U},\tr_B\circ\map{V}(\cdot\otimes\beta))\ge 1-\epsilon$.
\end{defi}
With these notions, the energy requirement of a unitary $\map{U}$ can be characterised by a lower bound on $E(\beta)$ of every $(E,\epsilon)$-ideal implementation of it, formulated in terms of $E$, $\epsilon$, and other properties of $\map{U}$. A relevant property is the energy constraint function that will be introduced next.
\subsection{Energy-constrained quantum channels}
In the last part of this section, we introduce the notion of energy constraint function and determine the family of unitary gates whose energy requirement is under consideration.
The energy requirement for a unitary that adds an arbitrarily large amount of energy to an input state with bounded energy is obviously infinite. Therefore, to make the setting meaningful, we consider energy constrained unitaries defined as follows: 
\begin{defi}[Energy-constrained unitary gates]
A unitary quantum channel $\map{U}$ acting on $\spc{H}$ is energy-constrained with an \emph{energy constraint function} $f:\R_+\to\R_+$, if 
\begin{align}
E(\map{U}(\rho))\le f\left(E(\rho)\right)
\end{align}
holds for every $\rho\in\St(\spc{H})$.
\end{defi}
The physical meaning of the energy constraint function $f(E)$ is the maximum output energy when the input energy is constrained to $E$. For example, in quantum optics it is common that channels are energy-constrained with linear constraint functions $f(E)=a E+b$ for some constants $a$ and $b$. Note that, by definition, $f(E)$ is non-decreasing. More properties of the energy constraint function can be found in the literature \cite{winter2017energy,Gschwendtner2021infinite}.

\section{Energy requirement of energy-limited unitary gates}\label{sec-mainresult}
\subsection{Battery recycling lemma}
In this subsection, we introduce a series of tools that will later be used to derive the energy requirement. The first tool is the battery recycling lemma, originally derived in \cite{chiribella2021fundamental} and extended to the unbounded Hamiltonian case in \cite{Gschwendtner2021infinite}. Compared to the battery recycling lemma in \cite{Gschwendtner2021infinite}, the version provided here is slightly more general, in that it includes unitary gates with generic energy constraint functions rather than assuming them to be linear. 

Intuitively, the lemma states that the battery state $\beta$ of any $(E,\epsilon)$-implementation of a unitary gate $\map{U}$ can be recycled for up to $O(1/\sqrt{\epsilon})$ times:
\begin{lemma}[Battery recycling; an energy-constrained version]\label{lemma-QIrecycling}
Let $(\map{V},\beta)$ be an $(E,\epsilon)$-ideal physical implementation of a unitary $\map{U}$. 
Suppose its inverse $\map{U}^{-1}$ has an energy constraint function $g(E)$. 
Then, there exists an energy non-increasing circuit $\map{N}:\spc{H}_S^{\otimes 2m}\otimes\spc{H}_B\to\spc{H}_S^{\otimes 2m}$ (consisting of multiple uses of $\map{V}$, $\map{V}^{-1}$ and $\tr_B$) such that
\begin{align}
D^{(E,\dots,E)}\left(\map{N}(\cdot\otimes\beta),(\map{U}\otimes\map{U}^{-1})^{\otimes m}\right)\le m\epsilon'
\end{align}
for any $m$, where $\epsilon'=\left(1+\frac{g(E)}{E}\right)\sqrt{\epsilon}$ and the energy constraints on all $2m$ input systems are equal to $E$.
\end{lemma}
Note that $\map{N}$ in the lemma is energy non-increasing, meaning that $E(\map{N}(\rho))\le E(\rho)$ for any $\rho$.
The content of the above lemma is slightly different from \cite[Lemma 11]{Gschwendtner2021infinite}, in that the  energy constraint function is allowed to be general, and the channel  $\map{N}$ emulates $m$ uses of $\map{U}\otimes\map{U}^{-1}$ rather than $\map{U}$.
The proof, however, can be derived in a similar way. For completeness of presentation, we provide the proof in \ref{app-proof-lemma}.

When $\|H_S\|<\infty$, the energy requirement can be obtained from the following argument:  $i)$ The battery recycling lemma implies that  the battery can be recycled to provide energy for up to $O(1/\sqrt{\epsilon})$ uses of $\map{U}\otimes\map{U}^{-1}$ up to a constant  $O(1)$ error. Hence,  the energy content of the battery has to be $O(1/\sqrt{\epsilon})$, up to {\it a correction term} due to the error of approximation. $ii)$ Then, using the Lipschitz continuity relation $|E(\rho)-E(\sigma)|\le \|H_S\|\cdot\|\rho-\sigma\|_1$ between the distance between two arbitrary quantum states $(\rho,\sigma)$ and their energy difference, we can upper bound the correction term and show that the energy requirement is indeed still $O(1/\sqrt{\epsilon})$.

The argument sketched in the previous paragraph is at the basis of the bound (\ref{eq-finite-bound}) in Ref.  \cite{chiribella2021fundamental}.  This approach, however, does not work for unbounded Hamiltonians. When $\|H_S\|=\infty$. There is no such continuity relation as required by step $ii)$. Consider, as a simple example, a Harmonic oscillator with $H_S=\sum_{k=0}^{\infty}k\hbar\omega|k\>\<k|$ (with $\omega>0$ being a constant) and two quantum states $\rho_0=|0\>\<0|$ and $\rho_1=\frac{1}{n} |n\>\<n|+(1-\frac{1}{n})|0\>\<0|$. respectively. It is obvious that the energy difference $E(\rho_1)-E(\rho_0)=\hbar\omega$ does not depend on the closeness $(1/2)\|\rho_0-\rho_1\|_1=\frac{1}{n}$ of the two states, which can be made arbitrarily small by increasing $n$. 

When $\|H_S\|=\infty$, to obtain an upper bound on the aforementioned correction term of the energy, we need a new upper bound
 on the \emph{minimum} energy of any state that is within a fixed distance to a certain state $\rho$.  
To this aim, we shall apply the following semidefinite program (SDP), which can be derived via first expressing the trace distance constraint in the SDP form \cite[Chapter 1]{watrous2018theory} and then applying duality. Compared to the existing Lipschitz continuity method, it offers a more accurate bound on the fluctuation of $E$ due to any disturbance to a state.
\begin{lemma}[Minimum energy SDP]\label{lemma-SDP}
Let $\rho\in\St(\spc{H})$ be a quantum state and denote by $H$ the Hamiltonian of the system $\spc{H}$. The minimum energy  achievable by states within the $\epsilon$-neighbourhood of $\rho$, $E_{\min,\epsilon}(\rho):=\min_{\sigma\in\St(\spc{H}):\frac12\|\sigma-\rho\|_1\le\epsilon}E(\sigma)$, is given by the following SDP (and its dual):
\begin{align}
E_{\min,\epsilon}(\rho)&=\begin{array}[t]{ll}\underset{A,X,Y}{\text{minimise}}\,\, & \tr(AH) \\
\text{subject to} \,\,&A\ge 0\quad \tr A = 1\quad \frac12(\tr X+\tr Y)\le 2\epsilon\\
&\left(\begin{matrix}X & -(\rho-A) \\ -(\rho-A) & Y\end{matrix}\right)\ge0.
\end{array}\\
&=\begin{array}[t]{ll}\underset{y,z,M}{\text{maximise}}\,\, & \tr\rho\left(\frac{z}2(M+M^\dag)-y\cdot I\right)-z\cdot2\epsilon \\
\text{subject to} \,\,&z\ge0\qquad M^\dag M\le I\qquad \frac{z}{2}(M+M^\dag)-y\cdot I\le H.\label{eq-SDP-dual}
\end{array}
\end{align}
\end{lemma}

An immediate observation is that, in Eq.~(\ref{eq-SDP-dual}), choosing a configuration of the parameters $(z,y,M)$ yields a lower bound of $E_{\min,\epsilon}(\rho)$. 
As an obvious example, we can always choose $z$, $y$, and $M$ so that $\frac{z}{2}(M+M^\dag)-y\cdot I\le H$ is saturated, which yields a lower bound $E_{\min,\epsilon}(\rho)\ge E(\rho)-2z\cdot\epsilon$. For finite systems, we can choose $z=\|H\|$ to get an effective bound $E_{\min,\epsilon}(\rho)\ge E(\rho)-2\epsilon\|H\|$. However, this bound becomes trivial for unbounded $H$. For unbounded $H$, we pick an \emph{energy threshold} $\overline{e}\ge 0$ and define the corresponding truncated Hamiltonian $H_{\rm trunc}=P_{\overline{e}}HP_{\overline{e}}$, where $P_{\overline{e}}$ is the projector onto 
the direct sum of all energy eigenspaces with energy
upper bounded by $\overline{e}$. Then, by choosing $(M+M^\dag)/2=H_{\rm trunc}/\overline{e}$, $z=\overline{e}$ and $y=0$ in Eq.~(\ref{eq-SDP-dual}), and defining the truncated energy to be 
\begin{align}\label{eq-E-trunc}
\overline{E}(\rho,\overline{e}):=\tr (P_{\overline{e}}HP_{\overline{e}}\rho),
\end{align} 
 we get:
\begin{cor}\label{cor-Emin-bound}
For any energy threshold $\overline{e}\ge0$, we have
\begin{align}
E_{\min,\epsilon}(\rho)\ge \overline{E}(\rho,\overline{e})-2\epsilon\cdot \overline{e}.
\end{align} 
\end{cor}

\begin{remark}
As a sanity check, let us apply Corollary \ref{cor-Emin-bound} to the state $\rho_1=\frac{1}{n} |n\>\<n|+(1-\frac1n)|0\>\<0|$ of a harmonic oscillator with $H_S=\sum_n n\hbar\omega|n\>\<n|$.  If the energy threshold is higher than $n\hbar\omega$, the energy is not truncated and the bound at best ($\overline{e}=n\hbar\omega$) reads $E_{\min,\epsilon}(\rho_1)\ge \hbar\omega(1-2\epsilon n)$. If the energy threshold is lower than $n\hbar\omega$, the energy of the truncated state is equal to $0$ and the bound becomes trivial. In both cases, the bound is non-positive if we choose $\epsilon=\frac{1}{n}$, which does not contradict with the fact that there exists another state $\rho_0=|0\>\<0|$ that is $\delta$-close but with zero energy.
\end{remark}

\subsection{Energy requirement for implementing unitary gates}

With all these preparations, we can now derive our main results on the energy requirement of implementing an arbitrary unitary that acts on a system with a potentially unbounded Hamiltonian. First, we present it in the most general form (Theorem \ref{thm-generalbound}) and then reduce it to an easier-to-use version (Theorem \ref{thm-single-bound}).
\begin{theo}[Energy requirement for implementing unitary gates]\label{thm-generalbound}
Consider any $(E,\epsilon)$-ideal physical implementation $(\map{V},\beta)$ of a unitary $\map{U}$ that acts on a system with a discrete Hamiltonian $H_S$ bounded from below. 
For any $m\in\N$, any energy threshold $\overline{e}^{(m)}$, and any $\rho_m\in\St(\spc{H}_S^{\otimes 2m})$ with $\tr(\rho H_k)\le E$ ($H_k:=I_{\overline{k}}\otimes (H)_k$ is the Hamiltonian of the $k$-th subsystem) for any $1\le k\le 2m$, the following inequality holds:
\begin{align}\label{eq-generalbound-m}
E(\beta)&\ge \overline{E}\left((\map{U}\otimes\map{U}^{-1})^{\otimes m}(\rho_m),\overline{e}^{(m)}\right)-2m\cdot E-2m\epsilon'\cdot \overline{e}^{(m)}
\end{align}
where $\overline{E}(\rho,\overline{e}^{(m)})$ is the truncated energy (\ref{eq-E-trunc}) and $\epsilon':=(1+g(E)/E)\sqrt{\epsilon}$ with $g(E)$ being the energy constraint function of $\map{U}^{-1}$.
\end{theo}
The new bound (\ref{eq-generalbound-m}), compared with \cite[Eq.~(1)]{chiribella2021fundamental}, is characteristic of an additional parameter $\overline{e}^{(m)}$, which controls the energy cut-off. This manifests the key difference between the energy bounded and unbounded cases, as the role of $\overline{e}^{(m)}$ is to address the issue of infinite $\|H_S\|$. We remark that $\overline{e}^{(m)}$, by definition, is a manually set variable independent of the system and the unitary to implement. In practice, nevertheless, we can often find an optimal cut-off that depends on the unitary to implement via simple algorithms (see Section \ref{sec-applications}).

\noindent{\em Proof of Theorem \ref{thm-generalbound}.}
For any $m\in\N$, consider any state $\rho_m\in\St(\spc{H}_S^{\otimes 2m})$ with bounded energy on every subsystem: $\tr(\rho H_k)\le E$ for any $1\le k\le 2m$.
Applying Lemma \ref{lemma-QIrecycling}, we know that there exists an energy non-increasing circuit $\map{N}$ that emulates $m$ uses of $\map{U}\otimes\map{U}^{-1}$ on $\rho_{m}$.
Since the network $\map{N}$ is energy non-increasing, the energy of the output state is bounded by the input energy, namely that
\begin{align}\label{eq-mainproof1}
E(\rho_{\rm out})\le E\left(\rho_m\otimes\beta\right) = E(\rho_m)+E(\beta).
\end{align}
Meanwhile, since the energy of $\rho_m$ is properly bounded, we can apply Lemma \ref{lemma-QIrecycling}, and thus the output state is $(m\epsilon')$-close to $(\map{U}\otimes\map{U}^{-1})^{\otimes m}(\rho_m)$, which implies that
\begin{align}
E(\rho_{\rm out})\ge E_{\min,m\epsilon'}\left((\map{U}\otimes\map{U}^{-1})^{\otimes m}(\rho_m)\right).
\end{align}
Further applying Corollary \ref{cor-Emin-bound} and combining with Eq.~(\ref{eq-mainproof1}) and $E(\rho_m)\le 2m\cdot E$ (by definition), we get Eq.~(\ref{eq-generalbound-m}).
\qed

In principle, we can choose $\rho_m$ to be an entangled state to maximise the bound. In practice, nevertheless, it is often convenient to pick a tensor-power form $\rho_m=\rho^{\otimes m}$ for some $\rho\in\St(\spc{H}_S^{\otimes 2})$.
If there exists an energy threshold $\overline{e}$ of the bipartite Hamiltonian $H_S\otimes I+I\otimes H_S$ such that $\overline{E}\left((\map{U}\otimes\map{U}^{-1})(\rho),\overline{e}\right)>E(\rho)$, then the general bound (\ref{eq-generalbound-m}) can be reduced to
\begin{align}
E(\beta)&\ge m\cdot \left(\overline{E}\left((\map{U}\otimes\map{U}^{-1})(\rho),\overline{e}\right)-E(\rho)-2m\epsilon'\cdot \overline{e}\right).
\end{align} 
Taking the maximum over $m\in\N$, we obtain the following specialised version of our main result:
\begin{theo}\label{thm-single-bound}
Consider any $(E,\epsilon)$-ideal physical implementation $(\map{V},\beta)$ of a unitary $\map{U}$ that acts on a system with a discrete Hamiltonian $H_S$ bounded from below.
For any energy threshold $\overline{e}$ of the bipartite Hamiltonian $H_S\otimes I+I\otimes H_S$ such that $\overline{E}\left((\map{U}\otimes\map{U}^{-1})(\rho),\overline{e}\right)>E(\rho)$ for some $\rho$ with $E(\rho)\le E$, the energy requirement satisfies the following bound:
\begin{align}
E(\beta)&\ge \frac{\left(\overline{E}\left((\map{U}\otimes\map{U}^{-1})(\rho),\overline{e}\right)-E\right)^2}{8\epsilon'\overline{e}}-\frac{\epsilon'\overline{e}}{2}.
\end{align}
Here $\epsilon':=(1+g(E)/E)\sqrt{\epsilon}$ with $g(E)$ being the energy constraint function of $\map{U}^{-1}$.
\end{theo}

\section{Applications of the energy requirement}\label{sec-applications}
\subsection{Systems with bounded Hamiltonian}
As the first example, we show how to retrieve the main result of Ref.~\cite{chiribella2021fundamental} on the energy requirement for energy-bounded systems.
When $\|H_S\|<\infty$, a legitimate truncation on $\spc{H}_S\otimes\spc{H}_S$ is the trivial one $\overline{e}=2\|H_S\|$ that keeps the Hamiltonian $H_S\otimes I+I\otimes H_S$ untouched, which yields $\overline{E}\left((\map{U}\otimes\map{U}^{-1})(\rho),2\|H_S\|\right)=E\left((\map{U}\otimes\map{U}^{-1})(\rho)\right)$. Substituting into Theorem \ref{thm-single-bound}, we get the following:
\begin{cor}[Energy requirement in the bounded Hamiltonian case]\label{cor-requirement-bounded}
The energy requirement of physically implementing a unitary $\map{U}$ (with the energy constraint fidelity $F^{E}\ge 1-\epsilon$) is bounded as:
\begin{align}
E(\beta)&\ge \frac{(\Delta E)^2}{16\epsilon'\|H_S\|}-\epsilon'\|H_S\|.
\end{align}
Here $\Delta E:=\max_{\rho}E\left((\map{U}\otimes\map{U}^{-1})(\rho)\right)-E(\rho)$ and $\epsilon':=(1+g(E)/E)\sqrt{\epsilon}$ with $g(E)$ being the energy constraint function of $\map{U}^{-1}$.
\end{cor}
Consider a generic bounded system Hamiltonian $H_S$ whose ground state energy is, without loss of generality, zero. Since the system has bounded energy, we can further waive the input energy constraint by letting $E=\|H_S\|$. Then, since $g(E)\le\|H_S\|$ we have $\epsilon'\le 2\sqrt{\epsilon}$. Moreover, we have $\Delta E=\max_\rho\tr\left(\Delta_U H_S\otimes I+I\otimes \Delta_{U^\dag} H_S\right)\rho$ with $\Delta_U H_S:=U^\dag H_S U-H_S$ is the change of $H_S$ upon the action of $\map{U}$. Observing that $\Delta_{U^\dag} H_S$ has the same spectrum as $-\Delta_U H_S$, we choose $\rho=\psi_1\otimes\psi_2$, where $\psi_1$ ($\psi_2$) is the eigenstate of $\Delta_U H_S$ corresponding to the maximal (minimal) eigenvalue $\lambda_{\max}(\Delta_U H_S)$ ($\lambda_{\min}(\Delta_U H_S)$). Therefore, Corollary \ref{cor-requirement-bounded} implies
\begin{align}
E(\beta)&\ge \frac{[(\lambda_{\max}-\lambda_{\min})(\Delta_U H_S)]^2}{32\sqrt{\epsilon}\|H_S\|}-2\sqrt{\epsilon}\|H_S\|,
\end{align}
matching the main result [Eq.~(1)] of Ref.~\cite{chiribella2021fundamental}.

\subsection{Displacement}
One major area of application for our new bound is quantum optics. As a working example, here we consider a Harmonic oscillator with Hamiltonian $H_S=\sum_{n=0}^{\infty}(n+\frac12)\hbar\omega|n\>\<n|$, where $\{|n\>\}$ is the photon (excitation) number basis. For simplicity, we follow the dimensionless convention and set $\hbar\omega=1$.
The unitary gate we consider is the displacement operator, one of the most fundamental building blocks of quantum optical circuits \cite{glauber1963coherent}. 

First, we determine the energy constraint function for a single-mode displacement operator $D(z)$ where $z\in\C$ is the displacement parameter. The Hamiltonian of the harmonic oscillator under consideration can be expressed as $H_S=(aa^\dag+a^\dag a)/2=(X^2+P^2)/2$, where $a$ ($a^\dag$) is the creation (annihilation) operator and $X$ ($P$) is the position (momentum) operator. 
$D(z)^\dag H_SD(z)=H_S+\sqrt{2}(X\Re(z)+P\Im(z))+|z|^2$, where $\Re$ ($\Im$) denotes the real (imaginary) part of a complex number, and we used the properties $D(z)^\dag XD(z)=X+\sqrt{2}\Re(z)$ and $D(z)^\dag PD(z)=P+\sqrt{2}\Im(z)$. Further using the Schwarz inequality $\<X\Re(z)+P\Im(z)\>\le \sqrt{\<X\>^2+\<P\>^2}\cdot|z|\le \sqrt{2\<H_S\>}\cdot|z|$, we get $\<D(z)^\dag H_SD(z)\>\le (\sqrt{\<H_S\>}+|z|)^2$, and a legitimate energy constraint function is thus
\begin{align}
f_{{\rm dis},z}(E)=\left(\sqrt{E}+|z|\right)^2\le 2(E+|z|^2).
\end{align}
Note that, since $D(z)^\dag=D(-z)$, the energy constraint function of $\map{D}(z)^{-1}$ is also $f_{{\rm dis},z}(E)$.

Choose the input state to be a coherent state $|\sqrt{E-\frac12}e^{i\arg(z)}\>\otimes |\sqrt{E-\frac12}e^{i\arg(-z)}\>$, which has energy $E$ on both modes. Sending it through $\map{D}(z)\otimes\map{D}(z)^{-1}$ yields $|\sqrt{E-\frac12}e^{i\arg(z)}+z\>\otimes |-(\sqrt{E-\frac12}e^{i\arg(z)}+z)\>$.
To obtain an energy requirement, we apply Theorem \ref{thm-single-bound} with some energy threshold $2\overline{e}>0$. Then the truncated output energy satisfies 
\begin{align}
\overline{E}_{\rm out}\ge & 2\overline{E}\left(\left|\sqrt{E-\frac12}e^{i\arg(z)}+z\right\>\left\<\sqrt{E-\frac12}e^{i\arg(z)}+z\right|,\overline{e}\right).
\end{align}
Since the energy distribution of the output state is Poisson, the truncated output energy can be bounded using the tail property of Poisson distributions. Defining $\nu:=(E+|z|^2+2|z|\sqrt{E-\frac12})/E>1$ to be the ratio between the output energy and the input energy, the truncated average energy can be bounded as
\begin{align}
\overline{E}_{\rm out}&=2\sum_{n\le\overline{e}-\frac12}\frac{(\nu E)^{n}\left(n+\frac12\right)e^{-\nu E}}{n!}\\
&\ge2\nu E\cdot\sum_{n\le\overline{e}-\frac32}\frac{(\nu E)^{n}e^{-\nu E}}{n!}.
\end{align}
Substituting into Theorem \ref{thm-single-bound} and noticing that $g(E)/E\le 2\nu$, we obtain the energy requirement of implementing a displacement $\map{D}(z)$: Define $\set{E}_{\rm thres}:=\left\{\overline{e}~:~\nu\cdot \sum_{n\le\overline{e}-\frac32}\frac{(\nu E)^{n}e^{-\nu E}}{n!}\ge1\right\}$ to be the set of legitimate $\overline{e}$. For any $(E,\epsilon)$-ideal implementation, the energy of the battery is lower bounded as:
\begin{align}\label{eq-EB-disp}
E(\beta)&\ge \max_{\overline{e}\in\set{E}_{\rm thres}}\frac{E^2}{2(1+2\nu)\overline{e}\sqrt{\epsilon}}\left(\nu\sum_{n\le\overline{e}-\frac32}\frac{(\nu E)^{n}e^{-\nu E}}{n!}-1\right)^2-\frac{(1+2\nu)\sqrt{\epsilon}\cdot\overline{e}}{2}.
\end{align}

 \begin{figure}[bht]
\centering
\subfigure[]{\label{fig-disp-E}
 \includegraphics[width=0.54\linewidth]{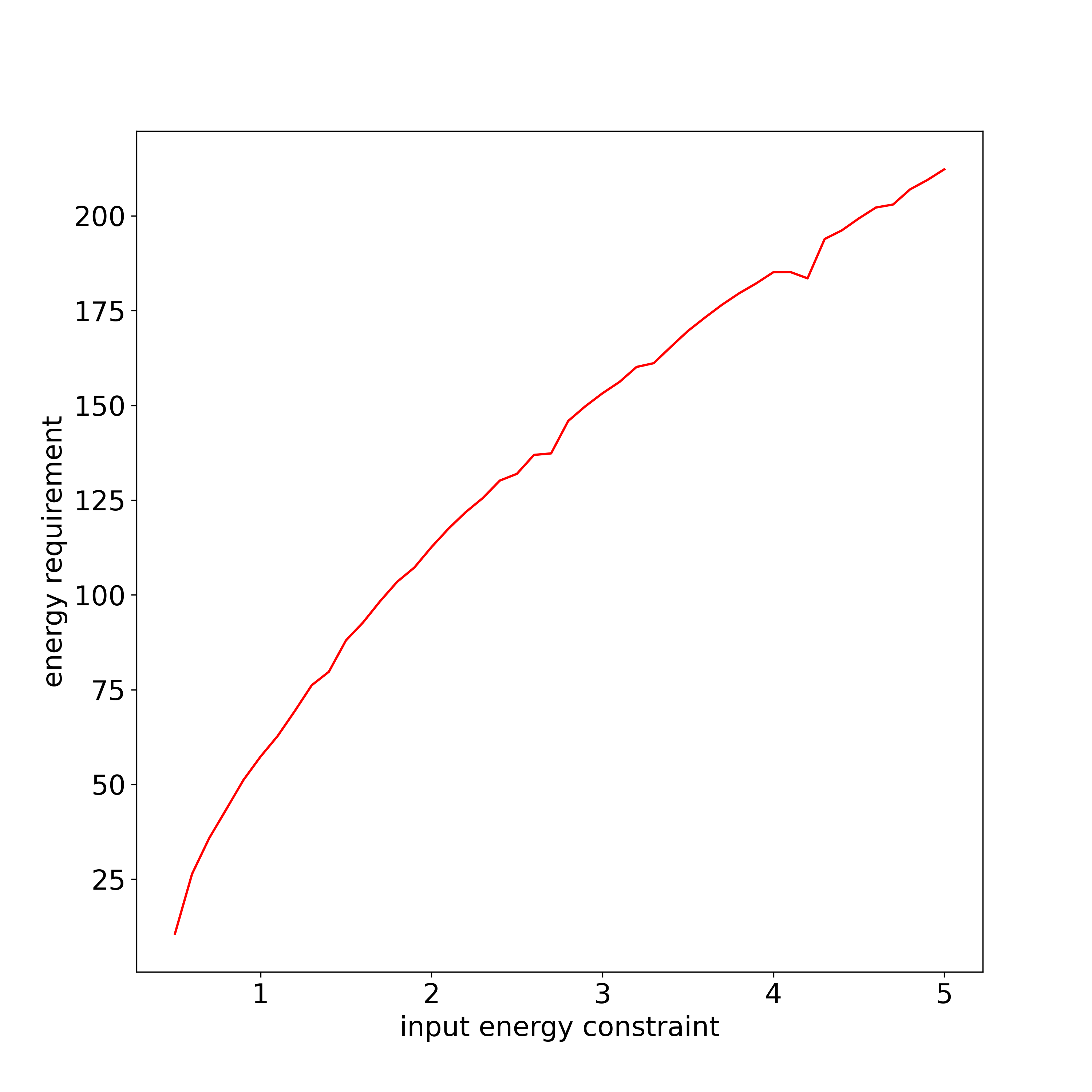}
}
\hfill
\subfigure[]{\label{fig-disp-err}
\includegraphics[width=0.41\linewidth]{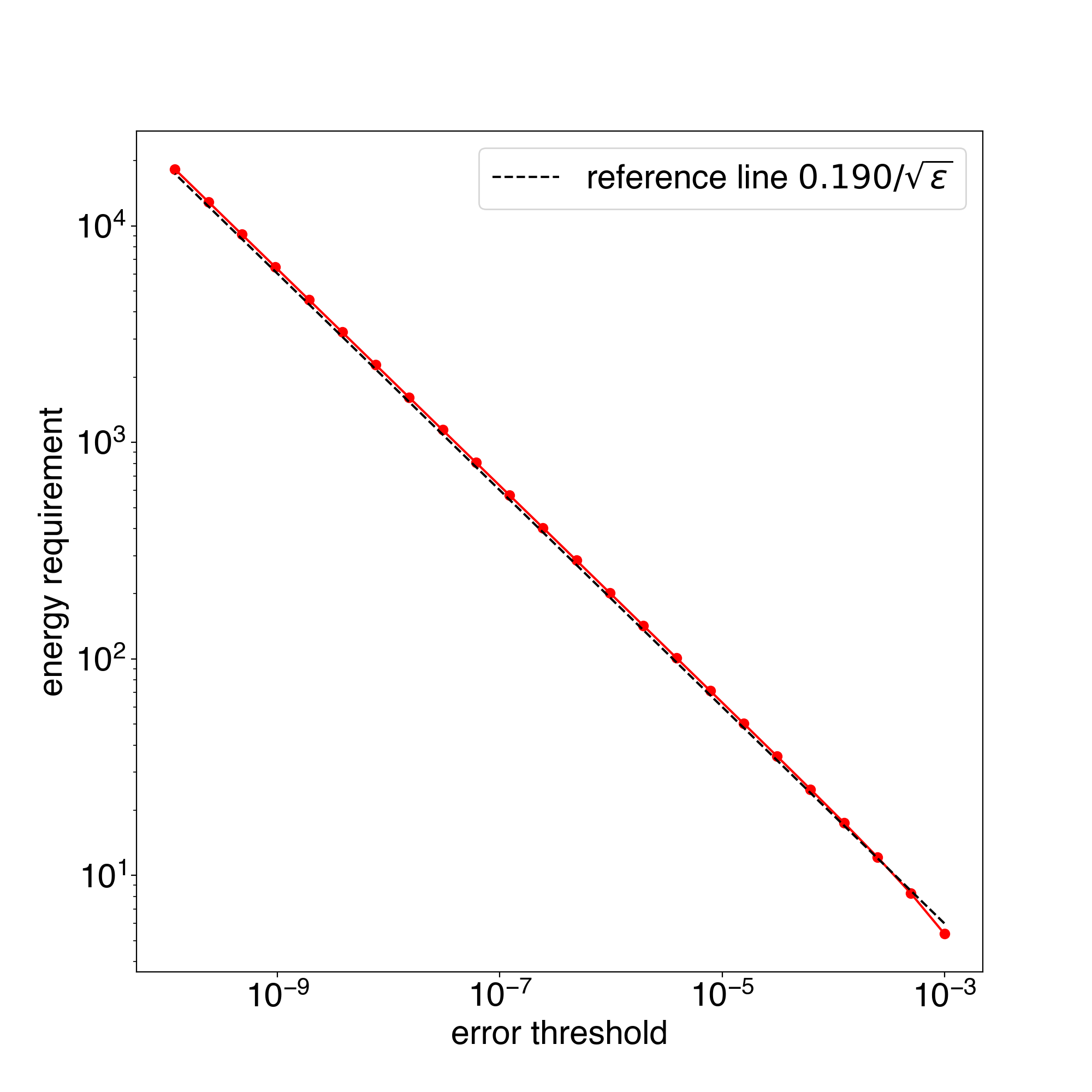}
}

\smallskip

\caption{  {\bf The energy requirement for a displacement $\map{D}(z)$ ($z=1$).} Figure \ref{fig-disp-E} shows the energy requirement (\ref{eq-EB-disp}) as a function of the input energy constraint $E$, whereas the error threshold is fixed to be $\epsilon=10^{-6}$. (The unit of the energy constraint is $\hbar\omega$.) It can be seen that the energy requirement increases with $E$, as the amount of energy that the displacement could generate also grows with $E$. Figure \ref{fig-disp-err} shows the log-scale plot of the energy requirement (\ref{eq-EB-disp})  as a function of the error threshold $\epsilon$, whereas the input energy constraint is fixed to be $E=4$. It can be seen that the energy requirement fits the $1/\sqrt{\epsilon}$ scaling except in the large $\epsilon$ region.
}
\label{fig-disp}
\end{figure}  

In Figure \ref{fig-disp}, the energy requirement for a displacement gate with displacement $z=1$ is plotted against the input energy constraint $E$ and the error threshold $\epsilon$. From Figure \ref{fig-disp-E}, we can see that the energy requirement indeed grows as the allowed input energy grows, since more energy has to be pumped into the battery to compensate the $E$-dependent energy generation. The plot suggests a sub-linear trend of the energy requirement growth. On the other hand, in Figure \ref{fig-disp-err} the energy requirement is plotted as a function of the error threshold $\epsilon$. We can see that the energy requirement is well-fitted by the curve $0.190/\sqrt{\epsilon}$ in the small $\epsilon$ region. The same phenomenon has been observed in the case of bounded Hamiltonians \cite{chiribella2021fundamental}.

At last, we discuss the behaviour of the bound in the vanishing error regime, i.e., $\epsilon\ll 1$. In this case, it is enough to choose the threshold $\overline{e}$ to grow (relatively slowly) with $1/\epsilon$. Following this intuition, we choose $\overline{e}=E\cdot\ln(1/\epsilon)$. Then, there exists an $\epsilon_0>0$ such that, when $\epsilon\le\epsilon_0$, we have $\sum_{n\le\overline{e}-\frac32}\frac{(\nu E)^{n}e^{-\nu E}}{n!}\ge\frac{\nu+1}{2\nu}$.  Substituting into Eq.~(\ref{eq-EB-disp}), we get 
\begin{align}
E(\beta)&\ge \frac{E}{\ln(1/\epsilon)\sqrt{\epsilon}}\cdot\left\{\frac{(\nu-1)^2}{9(1+2\nu)}-\frac{(1+2\nu)\epsilon(\ln(1/\epsilon))^2}{2}\right\}\qquad \text{for\ }\epsilon\le\epsilon_0
\end{align} 
where $\epsilon_0$ is a small enough positive constant. We can see that the energy requirement approximately achieves the same scaling (i.e. $1/\sqrt{\epsilon}$) as in the bounded Hamiltonian case \cite{chiribella2021fundamental}.

\subsection{One-mode squeezing}
Here we consider another fundamental unitary operation acting on a single harmonic oscillator: the single-mode squeezing operator \cite{yuen1976two}. The squeezing operator can generate entanglement between photons, enhancing the precision in the detection of extremely weak signals such as gravitational waves \cite{aasi2013enhanced}.

Again, we first determine the energy constraint function for a single-mode squeezing operator $S(\xi)$ where $\xi$ is the squeezing parameter. For simplicity, we assume $\xi>0$ to be a positive real number. $S(\xi)$ acts on the creation and annihilation operators as
$S^\dag(\xi)aS(\xi)=a\cosh\xi-a^\dag \sinh\xi$ and $S^\dag(\xi)a^\dag S(\xi)=a^\dag\cosh\xi-a \sinh\xi$. Therefore, the squeezing operator acts upon the Hamiltonian $H_S=(a^\dag a+aa^\dag)/2$ of a harmonic oscillator as $S^\dag(\xi)H_S S(\xi)=\cosh(2\xi)H_S-\frac{1}{2}\sinh(2\xi)((a^\dag)^2+a^2)$. Since $\<\frac12((a^\dag)^2+a^2)\>=\<H_S\>-\<P^2\>\ge-\<H_S\>$, the output energy can be bounded as $\<S^\dag(\xi)H_S S(\xi)\>\le (\cosh(\xi)+\sinh(\xi))^2\<H_S\>=e^{2\xi}\<H_S\>$, and thus a legitimate energy constraint function is
\begin{align}
f_{{\rm sq},\xi}(E)=e^{2\xi}E.
\end{align}
Note that, since $S(\xi)^\dag=S(-\xi)$, the energy constraint function of $\map{S}(\xi)^{-1}$ is also $f_{{\rm sq},\xi}(E)$. Therefore, we have $\epsilon'=(1+e^{2\xi})\sqrt{\epsilon}$ in Theorem \ref{thm-single-bound}.

For one-mode squeezing, we try different forms of input states and compare the obtained energy requirement bounds.
First, we choose the input state to be a number state $|\lfloor E-\frac12\rfloor\>^{\otimes 2}$, which has energy $\lfloor E-\frac12\rfloor+\frac12$ on both modes where $\lfloor\cdot\rfloor$ denotes the floor function. Sending it through $\map{S}(\xi)\otimes\map{S}(\xi)^{-1}$ yields two squeezed number states, whose photon number distribution is \cite{kim1989properties}
\begin{align}
P_{{\rm sn},\xi,E}(n)&=|\<n|S(\xi)|l\>|^2\quad\text{for}\quad l=\lfloor E-\frac12\rfloor\\
\<n|S(\xi)|l\>&=\frac{\sqrt{n!l!}}{(\cosh\xi)^{n+1/2}}\left(\frac{\tanh\xi}{2}\right)^{\frac{l-n}{2}}\cos^2\frac{(n-l)\pi}{2}\times S(\xi,l,n)\\
S(\xi,l,n)&:=\sum_{m=\frac{n-l}{2}}^{\frac{n}{2}}\frac{(-1)^m(2^{-1}\sinh\xi)^{2m}}{m!(n-2m)![m+(l-n)/2]!}
\end{align}

Alternatively, we choose the input state on each single mode to be a coherent state of energy $E$ with phase $\frac{\pi}{2}$. Sending it through $\map{S}(\xi)$  (and $\map{S}(\xi)^{-1}$ as well) yields a squeezed coherent states, whose photon number distribution is
\begin{align}
P_{{\rm sc},\xi,E}(n)=\left|\sum_{m=0}^{\infty}|\<n|S(\xi)|m\>\sqrt{\frac{\left(E-\frac12\right)^{m}e^{-(E-\frac12)}}{m!}}\right|^2.
\end{align} 
At last, we can also choose the input state to be a single-mode squeezed state of energy $E$ (i.e. with squeezing parameter $\frac12\cosh^{-1}(2E)$. Sending it over the gate $\map{S}(\xi)$ amounts to increasing its degree of squeezing by $\xi$. The output state is still a squeezed state with photon number distribution $P_{{\rm sq},\xi,E}(n)=|\<n|S(\xi+\frac12\cosh^{-1}(2E))|0\>|^2$.

Applying Theorem \ref{thm-single-bound} with some energy threshold $2\overline{e}>0$ (so that the threshold on each single system is at least $\overline{e}$), the truncated average output energy can be expressed as
\begin{align}
\overline{E}&=2\sum_{n\le\overline{e}-\frac12}\left(n+\frac12\right)P_{x,\xi,E}(n).
\end{align}
for $x=$ sn, sq, sc. By Theorem \ref{thm-single-bound}, we have
\begin{align}\label{eq-EB-sq}
E(\beta)&\ge \max_{\overline{e}}\frac{\left(\sum_{n\le\overline{e}-\frac12}\left(n+\frac12\right)P_{x,\xi,E}(n)-E_{\rm in}\right)^2}{2(1+\cosh(2\xi))\sqrt{\epsilon}\overline{e}}-\frac{(1+\cosh(2\xi))\sqrt{\epsilon}\overline{e}}{2},
\end{align}
where the maximisation is conducted over all $\overline{e}>0$ such that $\sum_{n\ge\overline{e}-\frac12}\left(n+\frac12\right)P_{x,\xi,\lfloor E-\frac12\rfloor}(n)$ is greater than the input energy $E_{\rm in}$, which is either $E$ ($x={\rm sc}, {\rm sq}$) or $\lfloor E-\frac12\rfloor+\frac12$ ($x={\rm sn}$).

 \begin{figure}[bht]
\subfigure[]{\label{fig-sq-err}
 \includegraphics[width=0.42\linewidth]{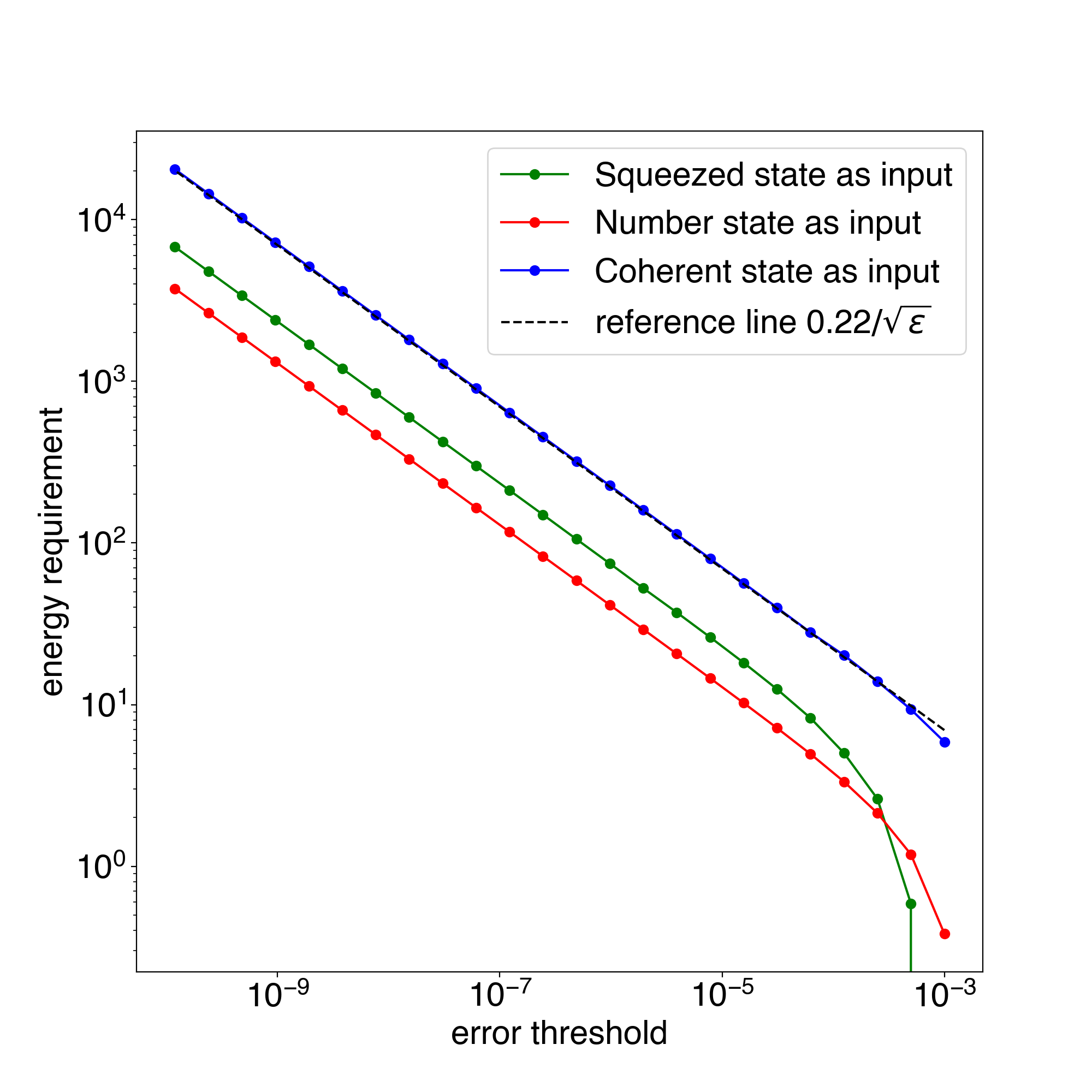}
}
\hfill
\subfigure[]{\label{fig-sq-dist}
\includegraphics[width=0.53\linewidth]{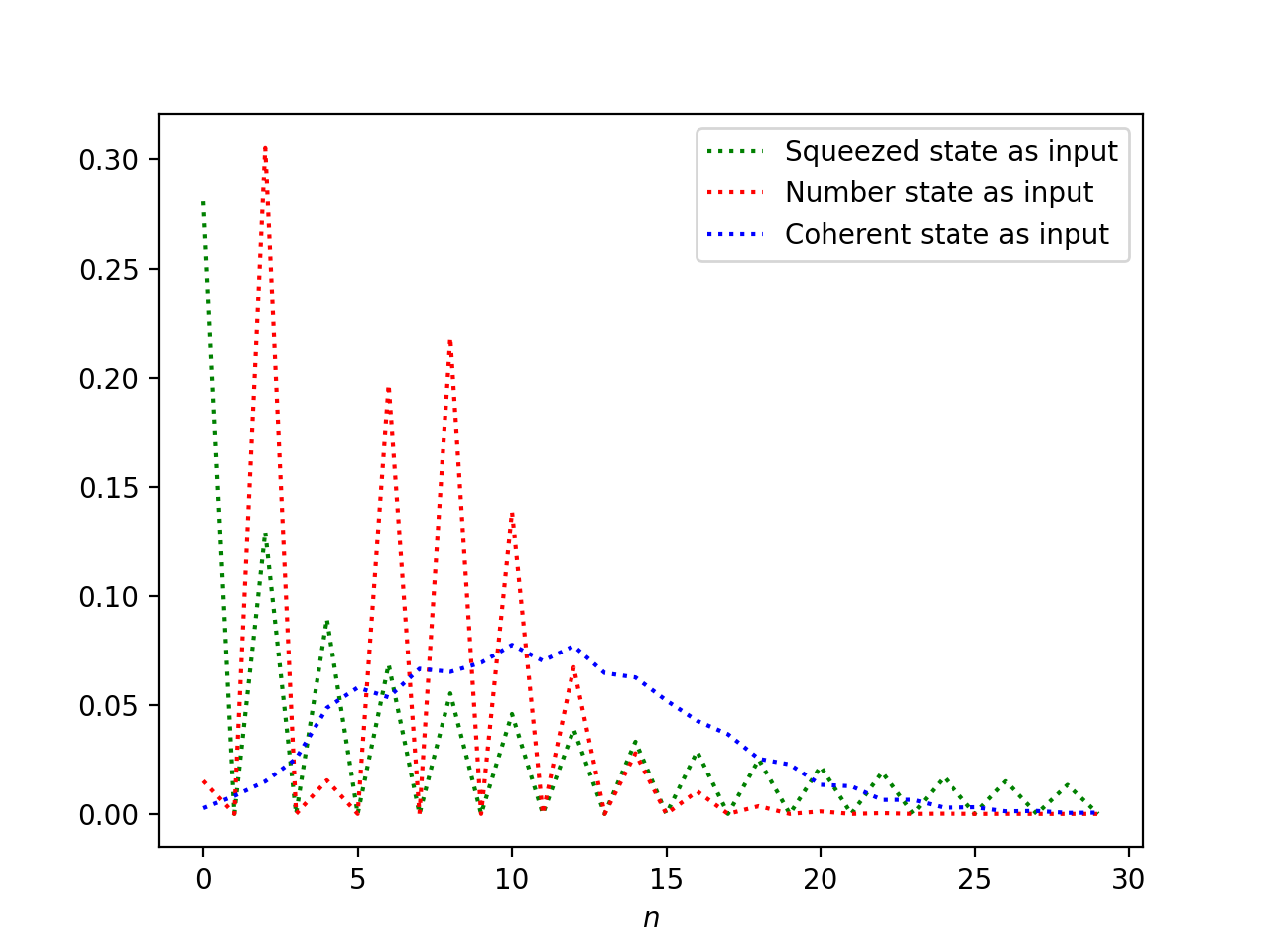}
}
\centering
\caption{  {\bf The energy requirement for a single-mode squeezing operation $\map{S}(\xi)$ ($\xi=0.5$) as a function of the error threshold $\epsilon$ for different choices of the input state.} In Figure \ref{fig-sq-err}, the energy requirement (\ref{eq-EB-sq}) is plotted with the input state being a squeezed state, a coherent state, or a number state. All choices of the input state have the same energy $E=4.5$. In Figure \ref{fig-sq-dist} shows the photon number distributions of the output state of $\map{S}(\xi)$ when choosing each of these states as the input state.
}
\label{fig-sq}
\end{figure}  

In Figure \ref{fig-sq-err}, the energy requirements obtained via inputting different states are compared. All input states have the same average energy $E=4.5$. By comparing with the reference line (black, dashed) we can see that the $(1/\sqrt{\epsilon})$-scaling persists.
The energy requirement obtained by plugging in a coherent state is the tightest, even though in this case the output energy before truncation ($=11.8$) is lower than that of the squeezed state case ($=12.2$). This seems to contradict the intuition that inputting a state that best stimulates the energy generating power of the unitary would make the bound tighter. The reason for such a phenomenon is manifested by Figure \ref{fig-sq-dist}, where the photon number distributions of the output states are plotted. It is clear from the plot that the tail of the output corresponding to the squeezed state case is longer than that of the coherent state case, leading to a larger value of $\overline{e}$ that decreases the value of the bound. Indeed, numerical calculations show that the optimal value of the energy threshold $\overline{e}$ is around 24 (which varies slightly as $\epsilon$ changes) in the coherent input state case in contrast to around 79 in the squeezed input state case. Meanwhile, although the optimal $\overline{e}$ in the number input state case is the smallest (around 15), the number state fails to trigger enough energy generation compared to the other two states. 
Therefore, a good choice of the input state (to make the bound tighter) should achieve a good balance between larger output energy and a more concentrated distribution with a shorter tail over the output energy spectrum. 

Finally, it is also intriguing to explore the relation between the above result and quantum metrology.
It was previously shown  that choosing \emph{the battery state} to be an optimal state for quantum metrology (specifically, for the estimation of the phase shift generated by the battery's Hamiltonian) will also achieve the optimal energy-precision tradeoff in the case of a bounded Hamiltonian \cite{chiribella2021fundamental}. 
Here we consider \emph{a different problem of choosing the input state} to test the accuracy of implementation.
Intuitively, a better state for quantum metrology as the input should provide a more stringent notion of accuracy, which could potentially lead to a higher energy requirement. In contrast, we showed that the squeezed state, a superior resource in photonic quantum metrology \cite{polino2020photonic}, did not perform better than the coherent state when chosen as the input state for evaluating our energy requirement bound. 
An immediate explanation, as mentioned in the above paragraph, is that there are other variables (such as the cut-off energy $\overline{e}$) than the accuracy threshold in our bound. 
Whether this phenomenon is fundamental, or there exist other bounds that lead to a different result, remains an interesting topic of future research.

\section{Conclusions and outlook}\label{sec-conclusion}
In this article, we discussed the energy requirement for implementing a unitary quantum gate. We derived a general lower bound on the amount of energy needed for the implementation, which extends the bound in Ref.~\cite{chiribella2021fundamental} to infinite dimensional systems and  unbounded Hamiltonians. To illustrate our new result, we analysed the energy requirement of operations  in quantum optics, such as  displacement operations and (single-mode) squeezing operations. The analysis can also be readily adapted to other operations, such as two-mode squeezing and non-Gaussian operations.

This article is reminiscent of Ref.~\cite{Gschwendtner2021infinite}, which generalises the optimal quantum programming result \cite{yang2020optimal} from finite dimensional systems to infinite dimensional systems. 
An important difference to our work is, however, that their resource of interest was the size (dimension) of the minimum quantum program, and the ``battery'' register of Ref.~\cite{Gschwendtner2021infinite} is thus constrained to be finite-dimensional by the task of programming. Here we completely replaced dimension constraints by energy constraints, which is more natural for many real physical systems such as a harmonic oscillator prepared in a coherent state.  

In Ref.~\cite{chiribella2021fundamental}, the method for determining the minimum resource requirement has also been extended from energy to general resources that satisfy three main assumptions, namely $i)$ monotonicity under discarding any subsystem, $ii)$ (sub)additivity on product states, and $iii)$ the Lipschitz continuity with respect to the trace distance between quantum states. In this article, despite the failure of $iii)$ (since the Lipschitz constant is infinite for unbounded Hamiltonians), we have successfully derived an energy requirement lower bound. This success prompts us to look for further relaxations of the assumptions in \cite{chiribella2021fundamental} and extend our work to other generic resources beyond energy.

Determining the resources that a unitary consumes is also relevant in the context of fundamental questions. A prominent example is the physics of black holes. A widespread assumption is that black holes, as viewed from the outside, can be treated as ordinary quantum systems. Under this assumption, the map that describes the evolution of a black hole together with the Hawking radiation that it produces would then be a unitary. There are furthermore strong indications that this unitary is rapidly mixing, and information-theoretic models of black holes thus usually rely on such a mixing property. For example, Page’s famous calculation of the time-dependence of the entropy of the Hawking radiation relies on the assumption that this unitary is typical, as if it was chosen at random according to the Haar measure from all possible unitaries~\cite{page1993information}. Similarly, Hayden and Preskill’s conclusions that an old black hole quickly emits all information that falls into it, and in this sense acts like a mirror, is based on the same typicality assumption~\cite{hayden2007black}. 

But if black holes are treated as quantum systems that evolve unitarily, it is reasonable to assume that they also obey the resource constraints that ordinary quantum systems do. Such considerations have already been made in terms of their complexity. It has been argued that the typicality assumption could be relaxed to the requirement that the unitaries are two-designs, which have a relatively low complexity. The work presented here suggests that it may also be interesting to ask whether energy considerations can constrain the class of physically plausible unitaries further. 

We note that black holes are usually considered as finite-dimensional quantum systems, and potentially one may apply the earlier results~\cite{chiribella2021fundamental} to their study. (Notice that the limitation of quantum information recovery in the Hayden-Preskill model of black holes has recently been considered in \cite{tajima2021universal}.) However, since this dimension is very large (actually as large as the dimension of a system can be that can still be embedded in spacetime without collapsing to a black hole), it may be more promising to consider energy rather than dimensional constraints. Furthermore, recent breakthrough results suggest that the subsystem structure of a spacetime containing a black hole is non-trivial~\cite{penington2020entanglement,almheiri2019entropy}. Based on these insights, it has been suggested to consider many-black-holes systems instead of a single black hole~\cite{renner2021black}. In such a system, it may be difficult to define the dimension of any individual black holes, so that energy bounds, again, appear to be a more promising choice. 

The application of our results to such a setting may however be subtle, not least because the notion of energy is depending on the reference one is considering. We thus leave it as a proposal for future work. 

\section*{Acknowledgements}

This work is supported by the Natural Science Foundation of Guangdong Province (Project 2022A1515010340), by HKU Seed Fund for Basic Research for New Staff via Project 202107185045,  by the Swiss National Science Foundation (SNSF) grant No. 200021\_188541, through the National Center for Competence in Research QSIT as well as through Grant No. 200021, and by the Hong Kong Research Grant Council (RGC) through Grants 17300918 and 17307520.  

\appendix 
\section{Proof of Lemma \ref{lemma-QIrecycling}}\label{app-proof-lemma}
By \cite[Proposition 1]{Shirokov2018} (which generalises the purification continuity \cite{kretschmann2008information} for the energy-constrained fidelity), there exists a pure state $\beta'$ such that $F^E(\map{V}\circ(\map{I}\otimes\beta),\map{U}\otimes\beta')\ge 1-\epsilon$. 
We remark that the purification continuity \cite{kretschmann2008information} and its extended form \cite{chiribella2013short,gutoski2017fidelity} has been used in a couple of previous works \cite{kretschmann2008complementarity,tajima2018uncertainty,takagi2020universal,tajima2020coherence,chiribella2021fundamental,tajima2021universal}.

By the Fuchs-Van de Graaf inequality \cite{fuchs1999cryptographic} of the energy-constrained channel fidelity, we have 
\begin{align}
D^E\left(\map{V}(\cdot\otimes\beta),\map{U}\otimes\beta'\right)\le \sqrt{\epsilon}.
\end{align}
By invariance of the energy-constrained error under the energy preserving unitary $\map{V}$, we have 
\begin{align}
D^E\left(\map{V}^{-1}\circ(\map{U}\otimes\beta'),(\cdot)\otimes\beta\right)\le \sqrt{\epsilon}.\label{eq-app-proof1}
\end{align}
Further applying properties of the energy-constrained diamond norm from \cite[Lemma 4]{winter2017energy}, we get:
\begin{align}
\sqrt{\epsilon}&\ge\frac{E}{g(E)}D^{g(E)}\left(\map{V}^{-1}\circ(\map{U}\otimes\beta'),(\cdot)\otimes\beta\right)\\
&\ge \frac{E}{g(E)}D^E\left(\map{V}^{-1}\circ(\map{U}\circ\map{U}^{-1}\otimes\beta'),\map{U}^{-1}\otimes\beta\right)\\
&=\frac{E}{g(E)}D^E\left(\map{V}^{-1}(\cdot\otimes\beta'),\map{U}^{-1}\otimes\beta\right).\label{eq-app-proof2}
\end{align}
Combining Eq.~(\ref{eq-app-proof1}) with Eq.~(\ref{eq-app-proof2}), we get 
\begin{align}
D^{(E,E)}\left((\map{V}^{-1}\otimes\map{I})\circ(\map{I}\otimes\map{V})(\cdot\otimes\beta\otimes\cdot),\map{U}^{-1}\otimes\beta\otimes\map{U}\right)\le  \left(1+\frac{g(E)}{E}\right)\sqrt{\epsilon}
\end{align}

Physically, the above inequality manifests the following fact: by consecutively applying $\map{V}$ and $\map{V}^{-1}$, each with its own system register, on the same battery $\beta$, up to an error of $\epsilon'$ one can simulate $\map{U}$ and $\map{U}^{-1}$ while keeping the battery ``untouched''. Iteratively applying the above procedure on the same battery register and new system registers for $m$ times, we get a network consisting of $\map{V}$ and $\map{V}^{-1}$ that acts on $2m$ identical subsystems $\spc{H}_{1},\dots,\spc{H}_{2m}\simeq\spc{H}_S$:
\begin{align}
\tilde{N}&=\map{V}_{2m}\circ\map{V}_{2m-1}\circ\cdots\circ\map{V}_1\\
\map{V}_k&:=\left\{\begin{matrix}\underbrace{\map{I}_{\overline{k}}}_{\text{all\ but\ }\spc{H}_{k}}\otimes\underbrace{\map{V}}_{\text{acting on }\spc{H}_{k}\otimes\spc{H}_B}\qquad &\text{for odd }k\\ &\\ \map{I}_{\overline{k}}\otimes\underbrace{\map{V}^{-1}}_{\text{acting on }\spc{H}_{k}\otimes\spc{H}_B}\qquad &\text{for even }k\end{matrix}\right.
\end{align}
At last, notice that discarding the battery register will not increase the energy since, by assumption, the energy eigenvalues of every register including the battery is non-negative.
Defining $\map{N}:=\tr_B\circ\tilde{\map{N}}$ to be the resultant network, we get the desired inequality
\begin{align}
D^{(E,\dots,E)}\left(\map{N},(\map{U}^{-1}\otimes\map{U})^{\otimes m}\right)\le m\left(1+\frac{g(E)}{E}\right)\sqrt{\epsilon}
\end{align}
thanks to the data processing inequality.\qed

\bibliography{ref} 
\bibliographystyle{unsrt} 

\end{document}